# Graphene Oxide Dielectric Permittivity at GHz and Its Applications for Wireless Humidity Sensing

[1]Xianjun Huang, [1]Ting Leng, [2]Thanasis Georgiou, [3]Jijo Abraham, [3,4]Rahul Raveendran Nair, [3,4]Kostya S. Novoselov, [1,4]Zhirun Hu[*]

[1]School of Electrical and Electronic Engineering, University of Manchester, Manchester, M13 9PL, UK. [2]BGT Materials Limited, Photon Science Institute, University of Manchester, Manchester, M13 9PL. [3]School of Physics and Astronomy, University of Manchester, Manchester, M13 9PL, UK. [4]National Graphene Institute, Manchester, M13 9PL, UK.

*Correspondence: z.hu@manchester.ac.uk

**Abstract:**

In this work, graphene oxide (GO) relative dielectric permittivity, both its real and imaginary parts, have been measured under various humidity conditions at GHz. It is demonstrated that the relative dielectric permittivity increases with increasing humidity due to water uptake. This finding is very different to that at a couple of MHz or lower frequency, where the relative dielectric permittivity increases with decreasing humidity. This GO electrical property was used to create a battery-free wireless radio-frequency identification (RFID) humidity sensor by coating printed graphene antenna with the GO layer. The resonance frequency as well as the backscattering phase of such GO/graphene antenna become sensitive to the surrounding humidity and can be detected by the RFID reader. This enables battery-free wireless monitoring of the local humidity with digital identification attached to any location or item and paves the way for low-cost efficient sensors for Internet of Things (IoTs) applications.

# Introduction

Graphene oxide (GO) is a chemical derivative of graphene functionalised with hydroxyl and epoxy groups. GO is a hydrophilic material and is capable to absorb a significant amount of water. Its water uptake depends strongly on the humidity of the environment and was previously studied by X-ray and neutron diffraction and in-situ electron microscopy. It was established that oxygen functional groups in GO drive intercalation of water molecules between individual GO sheets which results in an increase in the inter-layer spacing in GO thin films [1-5]. The presence of inter-layer water in a GO film can be crucial for a number of its properties, such as electrical conductivity [6, 7], molecular permeation [8, 9], mechanical [10] and dielectric properties [7, 11].

Multi-layered GO electrical and dielectric properties under various humidity conditions have been studied at low frequency [6, 7, 12]. Of particular interest would be the relative dielectric properties of multi-layered GO as a function of water uptake, considering the low intrinsic relative permittivity $\varepsilon_r$ of GO and the high $\epsilon_r$ of water at GHz. GO relative dielectric permittivity measurement at low frequency has been based on the equivalent circuit model of GO capacitor [6, 7,12]. While the equivalent circuit technique works well at low frequency, it is not suitable for high frequency, such as at GHz, due to parasitic effects. In this work the GO relative dielectric permittivity was obtained based on the measured transmission and reflection parameters (S-parameters) at GHz. There is no need of equivalent circuit model for the GO.

Pristine GO has been used in this work, which is a relatively good insulator at room temperature and low humidity. At high humidity, the ionic conductivity due to the intercalated water increases and GO becomes poorly conductive [6]. We experimentally determine both the real and imaginary parts of the GO relative dielectric permittivity at GHz. The findings in this work are very different to those reported in [7, 12] – both the real and imaginary parts of the GO relative dielectric permittivity decrease with decreasing humidity, from ~ 17 at 100% RH to 12 at 10% RH (the real part of the relative permittivity) and from 6 at 100% RH to 2 at 10% RH (the imaginary part of the relative permittivity). In particular, the imaginary part ($\varepsilon''$) changes by almost 200% depending on the water uptake, which is explained by a strong adsorption of RF waves by water. Furthermore, the electrical properties of the GO are used to construct battery-free wireless RFID humidity sensors for Internet of Things (IoTs) applications.

## Results and Discussions

**Extraction of GO relative dielectric permittivity under various humidity conditions through full electromagnetic wave simulation and experimental measurements.** The electrical property of GO can be completely characterized by its relative dielectric permittivity, $\varepsilon_r = \varepsilon' - i\varepsilon''$[6, 13]. There are several classical methods to measure relative permittivity in microwave band, including the transmission line (TL) method, free space method, resonator cavity, etc. [14]. However, all these methods do not suit permittivity measurement for small and thin piece of GO under different humidity environments.

Here, to measure the relative permittivity of the GO layer under various humidity conditions, a resonator circuit was designed (Fig. 1a,) with GO (thickness 30μm ± 2μm) printed on the top of the capacitor area (15mm×8mm) of the resonator (see Method for the details of GO preparation and sample fabrication). In order to extract the relative permittivity, a calibration circuit with exactly the same parameters was prepared, where GO layer was mimicked by a thin dielectric layer of exactly the same thickness as GO with known relative permittivity (see Supporting Materials, Fig. S1b).

Both the GO and the calibration circuits were placed in a hermetic container (2 litre in volume, see Supporting Materials, Fig. S3) in which constant humidity conditions were achieved by placing various saturated salt solutions inside the container. Three phase (vapour-liquid-solid) saturated salt solutions made of different salts were used to create different humid environments with constant RH values as these systems produce a constant vapour pressure over a long period of time [15,16]. The saturated salt solutions used were LiCl (RH-11%), $K_2CO_3$ (RH-43%), Mg $(NO_3)_2$ (RH-55%), NaCl (RH-75%) and $K_2SO_4$ (RH-98%) aqueous solutions prepared by dissolving excess amount of salts in deionised water. Before each measurement with each particular salt, the humidity was set to be stabilised for at least 48 hours. All measurements are done at 24°C. When the electrical property of GO (such as its permittivity) changes with humidity, it alters the loading of the resonator and results in a shift of the resonance frequency as well as change of the backscattering phase.

The measured transmission coefficients ($S_{21}$) for the samples with and without printed GO layer are displayed in Fig. 1b, together with the full electromagnetic wave simulation results for permittivity extraction (see Supporting Materials, Fig. S2). By comparing the simulated and measured transmission coefficients of the GO covered resonator, the GO relative dielectric permittivity under different humidity can be extracted.

From the measurement results, it becomes clear that the sample with GO layer has responded to the humidity change, whereas the sample without GO hasn't. The different responses of these two resonators can only be caused by the change of GO electrical properties due to water uptake. For the resonator with GO layer, it can be observed that the resonance shifts to lower frequency and its fractional bandwidth increases as the humidity rises. This reveals that both the real ($\varepsilon'$) and the imaginary ($\varepsilon''$) parts of the relative permittivity of GO increase as GO absorbs more water.

The resonance frequency of the GO covered resonator, as well as the extracted $\varepsilon'$, $\varepsilon''$ and the loss tangent ($tan\delta = \varepsilon''/\varepsilon'$) are presented in Fig. 2a,b. It can be seen that $\varepsilon'$ and $\varepsilon''$ of the GO change from about 11 to 17.6 and 2.3 to 6.4, respectively, as RH varies from 11% to 98%. These findings are very different to those published works at low frequency. It is revealed that GO permittivity can be very different in different frequencies. At low frequency, a large permittivity change can be observed [7, 12], whereas the change is much smaller at high frequency. This is probably due to the orientation polarization of absorbed water. At low frequency, the polarization of the water can follow the electrical field direction and hence large permittivity changes as humidity varies. At high frequency, the electrical field direction changes fast so that the polarization of the water can't catch up and hence the dielectric permittivity has relatively smaller change with humidity [17]. Water has dielectric permittivity of ~ 80. As humidity increases, more water will be absorbed by the GO hence higher permittivity [17].

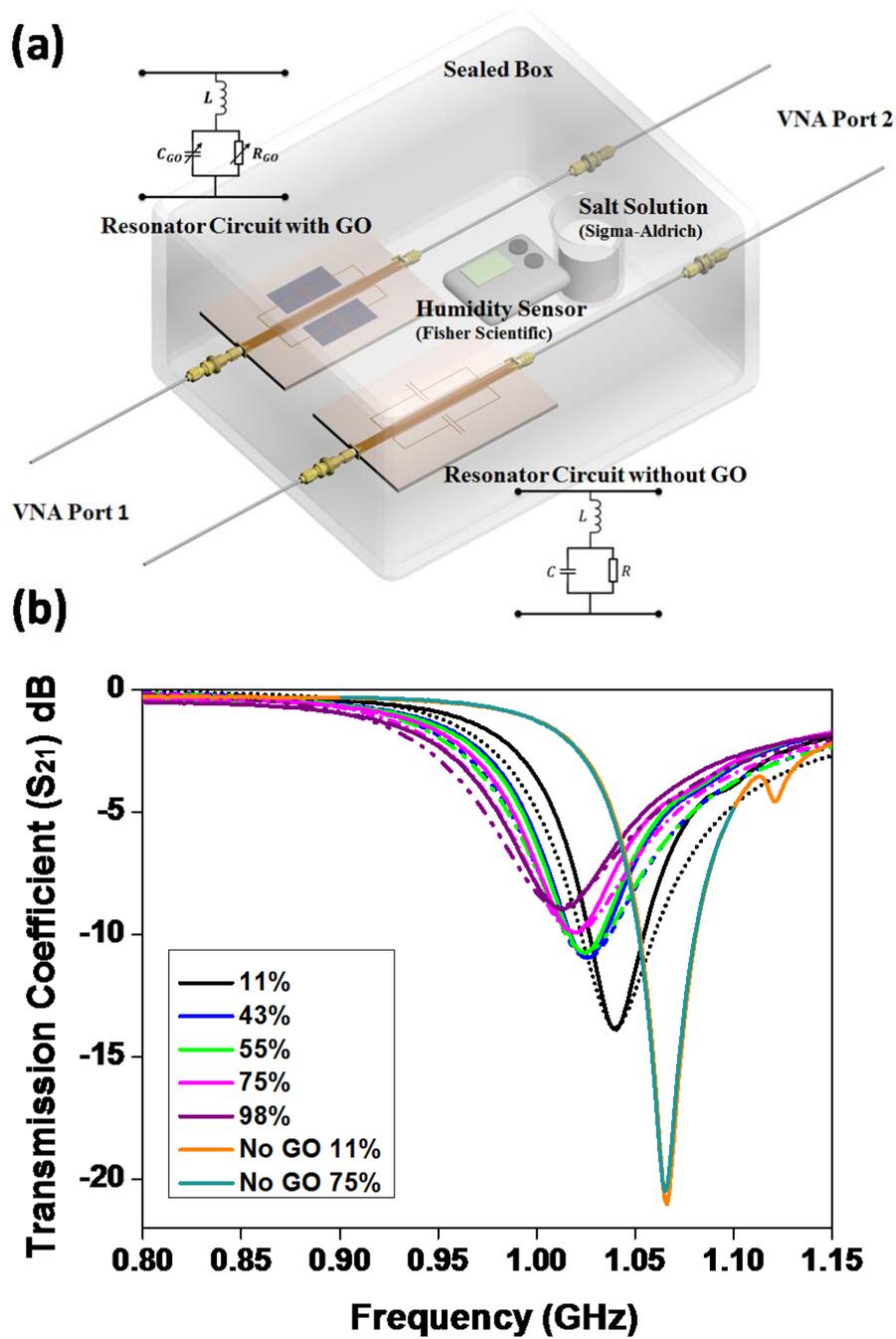

Figure 1: (a) Resonator circuit for GO permittivity measurement and (b) Measured (solid lines) and simulated (dashed lines) transmission coefficients ($S_{21}$) of the samples with/without GO layer for various RH.

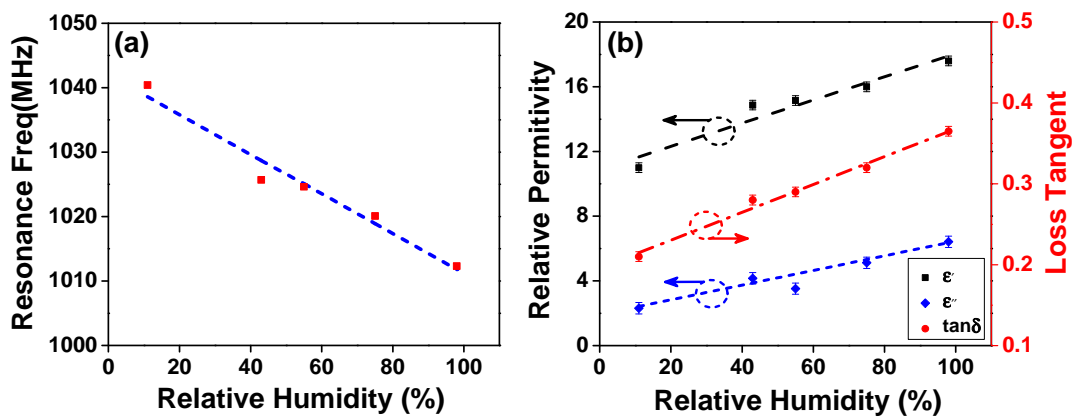

Figure 2: (a) Resonance frequency as function of RH and (b) Relative permittivity components and the loss tangent of the GO under various humidity conditions.

**Battery-free Wireless GO sensing enabled by printed graphene RFID technology.** It's well known that GO is sensitive to humidity [7, 12]. However, the sensing mechanism proposed here is different to those published works. In this work, the GO layer was directly coated on the graphene radio-frequency identification (RFID) antenna. Instead of using GO capacitor to sense the humidity [7, 12], the phase shift of backscattering signal due to the humidity change was detected by the RFID reader. The GO sensor is battery-free, wireless and fully printable. Battery-free wireless sensing is in the heart of IoTs technology [18, 19], allowing collection of information about the immediate state of the object without the need of batteries. Below we demonstrate a battery-free RFID humidity sensor by combining printable graphene RFID antenna with GO coating.

RFID antennas are sensitive to changes in the environment due to proximity effects. When a layer of GO is printed on top of an RFID antenna – the resonance frequency of the latter become sensitive to the permittivity of GO layer (and thus to humidity) as it alters the antenna impedance. Other parameters, such as the phase of the backscattered signal, change as well, which can be easily detected by the RFID reader. The operating principle and use of the phase of the backscattered signal to measure the relative humidity (RH) with the GO-coated printed graphene RFID antenna, shown in Fig. 3b, are explained and demonstrated in this work.

When a RFID reader transmits an electromagnetic wave signal (also called 'forward electromagnetic wave signal') to an RFID antenna, the antenna draws energy from this forward signal and activates the RFID chip on the antenna. The backscattered signal is both amplitude and phase modulated by the RFID chip through varying the chip's input impedance. Modulation occurs as the RFID chip rapidly switches between two discrete impedance states [20, 21].

The operating principle and an equivalent circuit for the antenna's amplitude and phase modulation are schematically shown in Fig. 3 (detailed information about the experimental setup for sensing measurement can be found in Supporting Materials, Fig. S4).

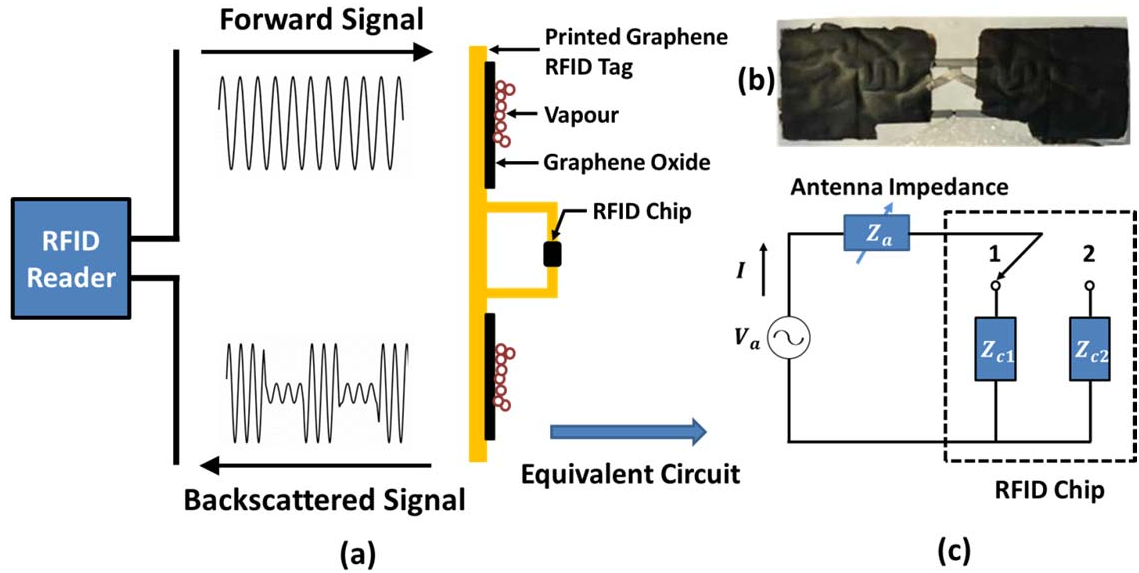

Figure 3 : (a) Operating principle of the GO based printed graphene RFID sensor system. (b) Printed graphene antenna with a layer of GO on top. The thickness of GO layer is 15 μm. (c) The equivalent circuit of and RFID tag.

Different to a regular antenna, impedance of a RFID antenna is typically designed to conjugately match to the higher impedance state of the chip in order to maximize the collected power. The equivalent open source voltage $V_a$ on the antenna in Fig. 3c can be given as [22]:

$$V_a = \sqrt{8P_{Ant}Re(Z_a)}, \qquad (1)$$

where $P_{Ant}$ is the power available at the antenna port, $Z_a$ is the antenna impedance. The switching between the two input impedance states $Z_{C1}$ and $Z_{C2}$ generates two different currents at the antenna port, which can be calculated as [22]:

$$I_1 = V_a \left(\frac{1}{Z_a+Z_{c1}}\right) \qquad (2)$$

$$I_2 = V_a \left(\frac{1}{Z_a + Z_{c2}}\right) \qquad (3)$$

When the humidity changes, the GO layer on RFID antenna changes its dielectric property. At high humidity, the ionic conductivity due to the intercalated water increases and even pristine GO becomes conductive but only poorly (mega ohms resistance at 100% RH and Giga ohms at 0% RH [23]). The resistance of the GO coating is still several orders of magnitude higher than the resistance of the printed graphene RFID antenna in this experiment (fractions of Ohm [24]), so the effects of the change of the GO conductivity can be ignored and only the change in dielectric property are taken into account. The GO dielectric property change alters the antenna impedance $Z_a$. As $Z_a$ changes so do $I_1$ and $I_2$, causing the backscattered signal phase varies accordingly. The backscattered signal phase change can be detected by the RFID reader. In this work, the backscattered signal phase was measured using Voyantic Tagformance under various humidity conditions and depicted in Fig. 4 [25].

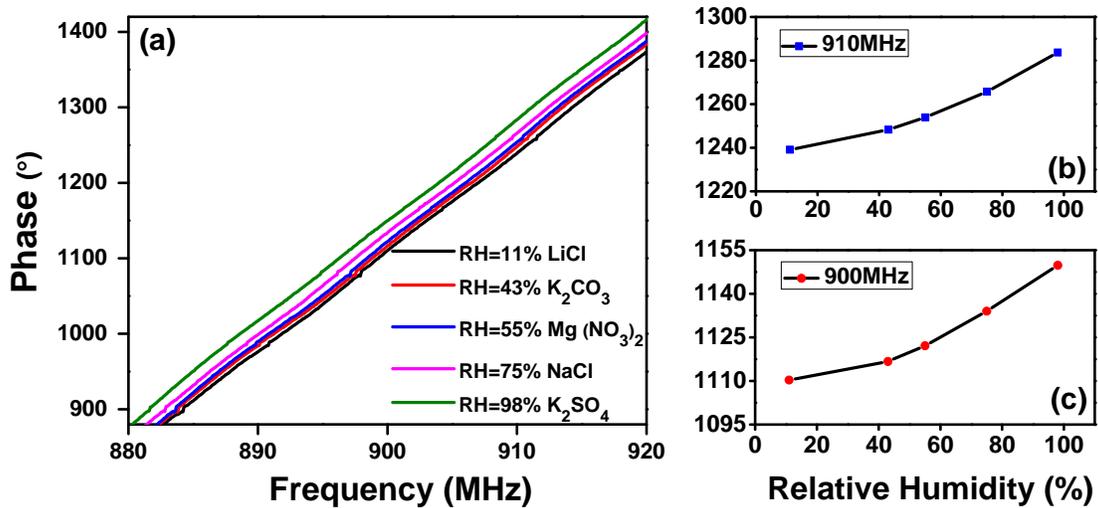

Figure 4: (a) Measured backscattered signal phases with various humidity as functions of frequency, (b) enlarged backscattered signal phases at 910 MHz as function of humidity and (c) enlarged backscattered signal phase at 900 MHz as function of humidity.

From Fig. 4a, it can be seen that the humidity has clear effects on the backscattered signal phase at typical RFID frequency spectrum from 880 MHz to 920 MHz, which experimentally proves that the backscattered signal contains humidity information. Together with the ID information of the sensing tag, a printed graphene enabled battery-free RFID GO humidity sensing system is presented. As it can be seen from Fig. 4b and c, the backscattered 910 MHz and 900 MHz signal phases increases by 44.6° and 39.5°, respectively, as RH rises from 11% to 98%. For 910 MHz signal, average phase change of 0.5° every 1% RH (sensitivity of

0.5°/1% *RH*) can be observed. Unambiguously demonstrating the effectiveness of wireless printed graphene enabled battery-free RFID GO humidity detection.

It is worth noticed that the technique used here to detect the humidity change is very different to that employed in other reported printed battery-free UHF RFID sensors [26-28]. In those reported works, the minimum power-on-tag was measured and the resonance frequency was then extracted from the minimum power-on-tag. This technique requires the reader to scan the whole allocated UHF RFID frequency spectrum and after-measurement data process to find out the minimum power-on-tag and resonance frequency. In this work, the backscattered signal phase was measured. The advantage of measuring backscattered signal phase is that there is no need for the reader to scan the whole allocated frequency spectrum in order to find out the resonance frequency. As it can be seen in Fig. 4, the humidity change can be simply detected at a single frequency point, which greatly simplifies and speeds up the measurement.

## Conclusions

We have experimentally extracted the GO relative dielectric permittivity under various humidity conditions at GHz. The measurement results clearly reveal that the GO dielectric property (relative permittivity, or dielectric constant and loss tangent) changes with the humidity but in a different manner as it does in a couple of MHz or lower frequency. Most distinguishingly, the relative dielectric permittivity does not have large changes (from ~ten to a few thousands [7]) and decreases with decreasing humidity at GHz. Furthermore the dielectric property has been used to design and build a RFID sensing tag which can act as a battery-free wireless humidity sensor, by coating GO layer on top of the printed graphene RFID antenna. Such combination can form bases for future energy harvesting enabled RFID sensors for IoTs applications. Furthermore, backscattered signal phase rather than minimum power-on-tag or resonance frequency has been used to detect the humidity change, which can significantly simplify and speed up the monitoring process.

## Method

**Full Electromagnetic Wave Simulation:** Commercially available CST MICROWAVE STUDIO 2015 is used for the full electromagnetic wave simulation. CST can solve Maxwell's equations numerically in both time and frequency domains. The resonators coated with GO and without GO are simulated. Waveguide ports are used to feed the simulated structures. The ports are matched to the ports of the structures to excite the fundamental propagation mode and to ensure a low level of reflection. CST can provide many outputs based on the simulation results calculated from Maxwell's equations, such as electric field, magnetic field, Poynting vector, scattering parameters (S-parameters), etc. The S-parameters were used to extract the relative dielectric permittivity in this work.

**Preparation of GO:** Modified Hummers method was employed to prepare GO. The typical oxygen content for GO produced by this technique is around~ 30-40% [29,30]. In brief, 4 grams of graphite was mixed with 2 grams of $NaNO_3$ and 92mL of $H_2SO_4$. $KMNO_4$ was subsequently added in incremental steps in order to achieve a homogeneous solution. The temperature of the reaction was monitored and kept near 100$^o$C. The mixture was then diluted by 500mL of deionised water and 3% $H_2O_2$. The resulting solution was washed by repeated centrifugation until the pH value of the solution was around 7. The GO was then diluted to the required concentration. Lateral size of GO flakes is about 500 µm× 500µm.

The water uptake with two different flake sizes (0.5 µm and 10 µm) has been measured by monitoring the weight change of GO exposed to different humidity conditions. The results show that the small and large flakes only have a few percentage decrease in mass uptakes indicating similar hydration behaviour (see Supporting Materials, Fig. S5). The interlayer spacing measurement using X-Ray Diffraction (XRD) is consistent with mass uptake data showing monotonic increase of interlayer spacing of GO from 6.5 Å to 10 Å by changing humidity from 0 to 100% [31].

For the purpose of coating GO on a printed graphene RFID antenna, a 10 grams per litre viscous GO solution was used. This allowed direct screen printing of the GO on the antenna, which was left to dry overnight in a fume hood under continuous air flow. The printed graphene RFID antenna is made with screen printing and rolling compression [24,32]. The lateral SEM view of the GO coated on printed graphene on paper substrate is shown in Fig. 5. As it can be seen, the three-layer structure is obvious and clear - GO layer, printed and compressed graphene layer and paper substrate, stacked in sequence from top to bottom.

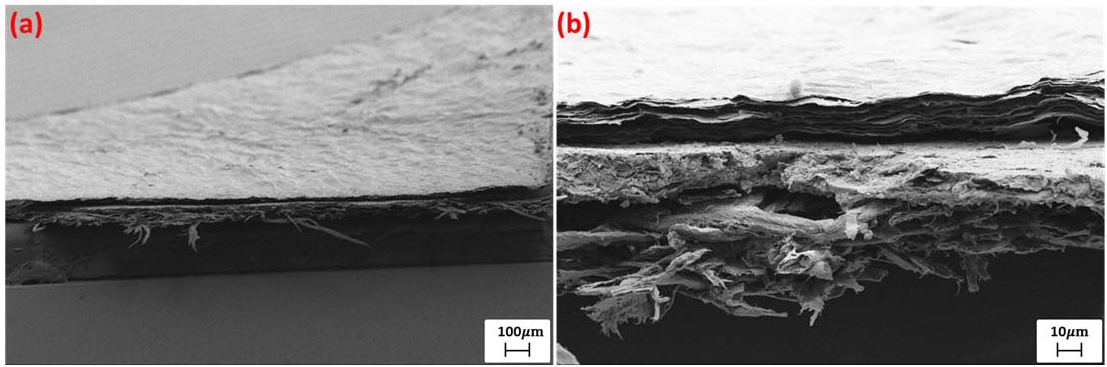

Figure 5: SEM view of GO layer on printed graphene RFID antenna on paper substrate. (a) Large view; (b) Enlarged view, layers from top to bottom are GO, printed graphene and paper in sequence.

**Acknowledgement:** The project is partially supported by UK EPSRC under EPN010345. Thanks also to BGTM Ltd for printing the graphene antenna pattern.

**Author Contributions:**

Xianjun Huang designed and prepared the experiments, measured and analysed the experimental data and drafted the manuscript. Ting Leng participated in measurements and drafted the manuscript. Jijo Abraham prepared salt solutions for the humidity measurement. Rahul Raveendran Nair advised GO permittivity measurement setup and partly drafted manuscript. Konstantin Novoselov initiated and coordinated the project as well as drafted the manuscript. Thanasis Georgiou prepared and coated GO. Zhirun Hu initiated and supervised the project as well as drafted the manuscript. All authors have given approval to the final version of the manuscript.

**Additional information**

Competing financial interests: The authors declare no competing financial interests.

# Supplementary Information

## Graphene Oxide Dielectric Permittivity at GHz and Its Application for Wireless Humidity Sensing

[1]Xianjun Huang, [1]Ting Leng, [2]Thanasis Georgiou, [3]Jijo Abraham, [3,4]Rahul Raveendran Nair, [3,4]Kostya S. Novoselov, [1,4]Zhirun Hu*

*Correspondence: Zhirun Hu (z.hu@manchester.ac.uk)

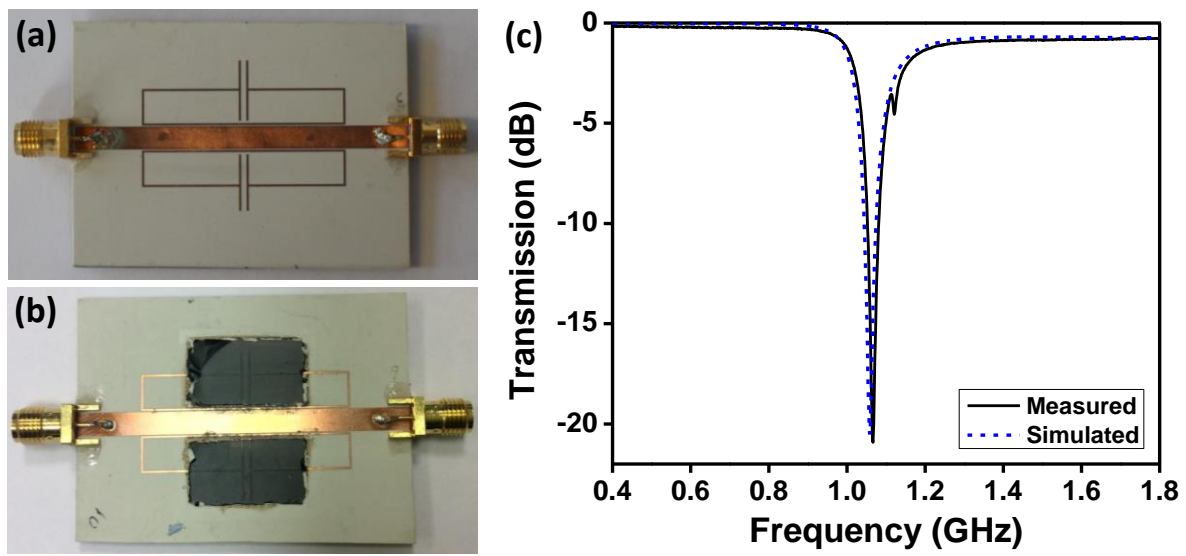

Fig. S1. (a) Microstrip resonator without printed GO layer, (b) Microstrip resonator with printed GO layer ($15\ mm \times 8\ mm$). The thickness the GO is 30 μm and (c) Simulated and measured transmission coefficients ($S_{21}$).

Fig. S1 shows the designed resonator for GO permittivity measurement and extraction. To validate the full electromagnetic wave simulation (CST Microwave Studio), we have compared the simulated and measured transmission coefficients of the resonator without GO layer, Fig. S1 (c). It can be seen that the simulated results agree very well with the measured ones, validating the simulation.

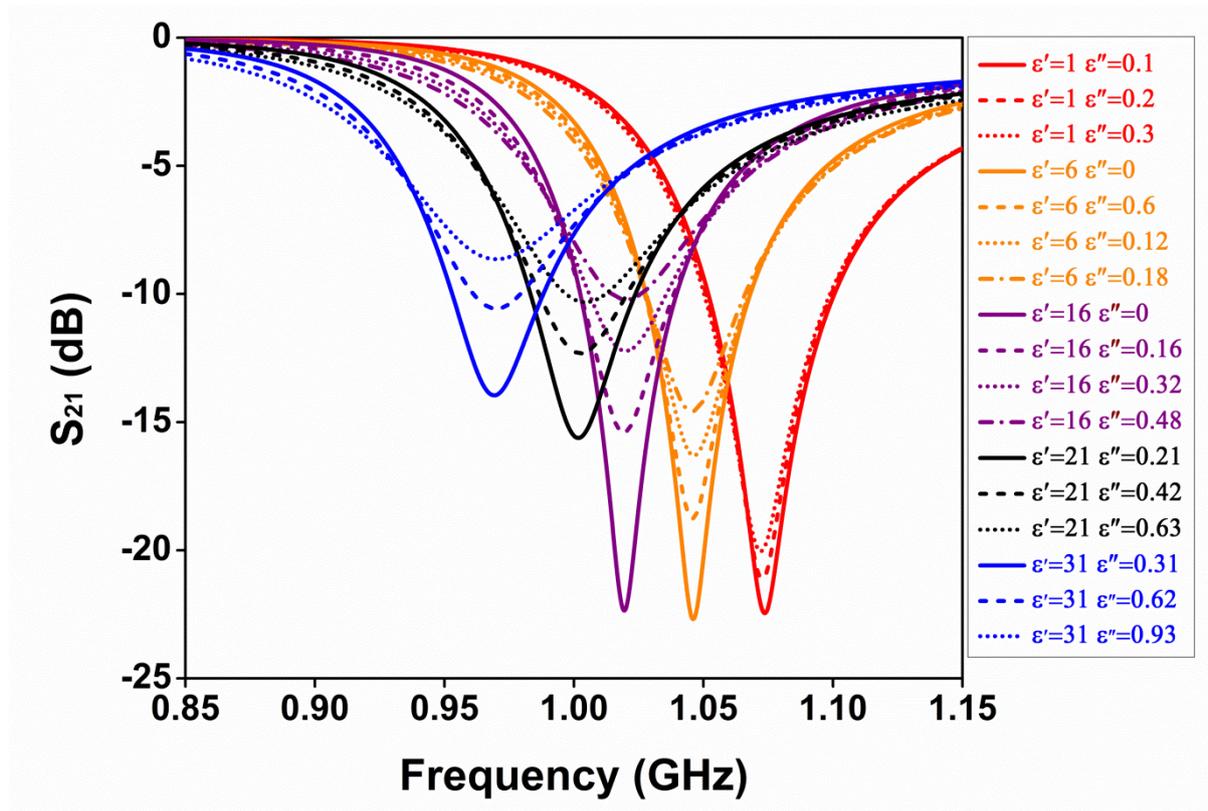

Fig. S2. Simulated transmission coefficients ($S_{21}$) of the resonator with dielectric layer of various relative permittivity ($\varepsilon'$ and $\varepsilon''$).

To extract the relative permittivity of the GO under various humidity conditions, GO was mimicked by thin dielectric layer, which has exactly the same size, thickness and location as that shown in Fig. S1 (b). The transmission coefficients of the resonator of 5 different sets of relative permittivity ($\varepsilon_r = \varepsilon' - i\varepsilon''$) were simulated and shown in Fig. S2. Each colour set of the curves in the figure contains the same real part ($\varepsilon'$) but various imaginary part ($\varepsilon''$) of the relative permittivity. It can be observed that for the same $\varepsilon'$, the resonance frequency changes little with $\varepsilon''$. This is because $\varepsilon''$, which is related to material loss tangent ($tan\delta = \varepsilon''/\varepsilon'$), mainly affects the Q factor of the resonator. The simulations reveal that the changes of relative permittivity pose obvious alteration on the resonator's transmission performance. GO permittivity can be extracted by comparing the experimental measurements and full electromagnetic wave simulations.

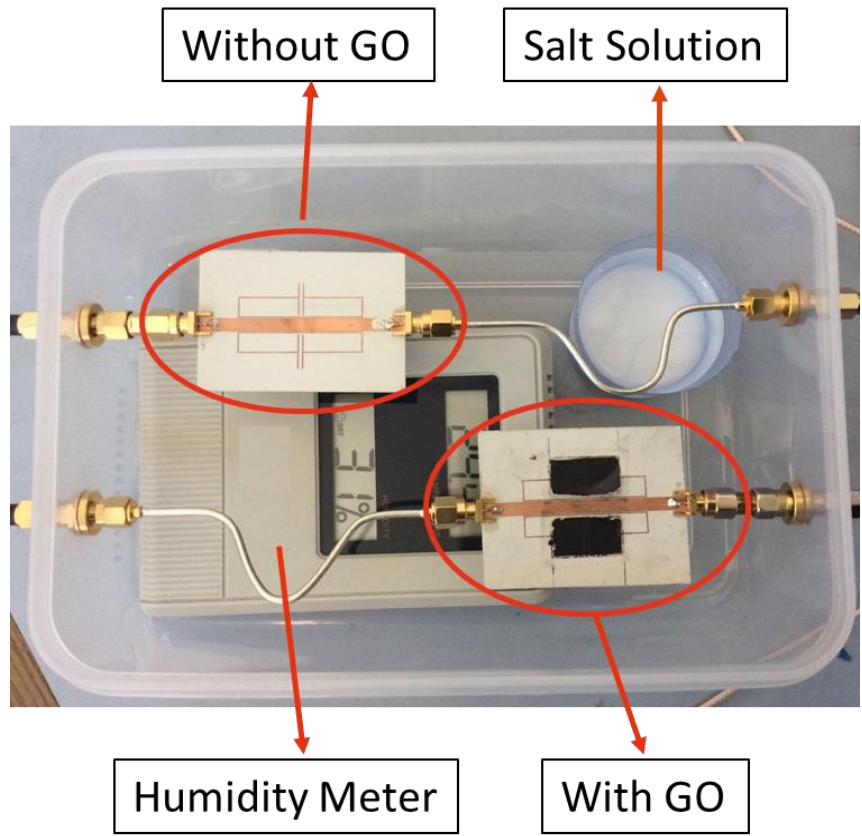

Fig. S3. Sealed box setup for GO permittivity measurement (the top cover of the box has been removed for a better view).

Fig. S3 shows the GO permittivity measurement setup. A digital humidity meter (Fisher Scientific 116617D) is placed in the box for monitoring. Rubber-tight SMA connectors (RS Stock No.716-4798) are used to connect inside/outside microwave cables, making the box well sealed. The out-extended cables are connected to VNA (Agilent E5071B) for scattering parameter (S-parameter) measurements.

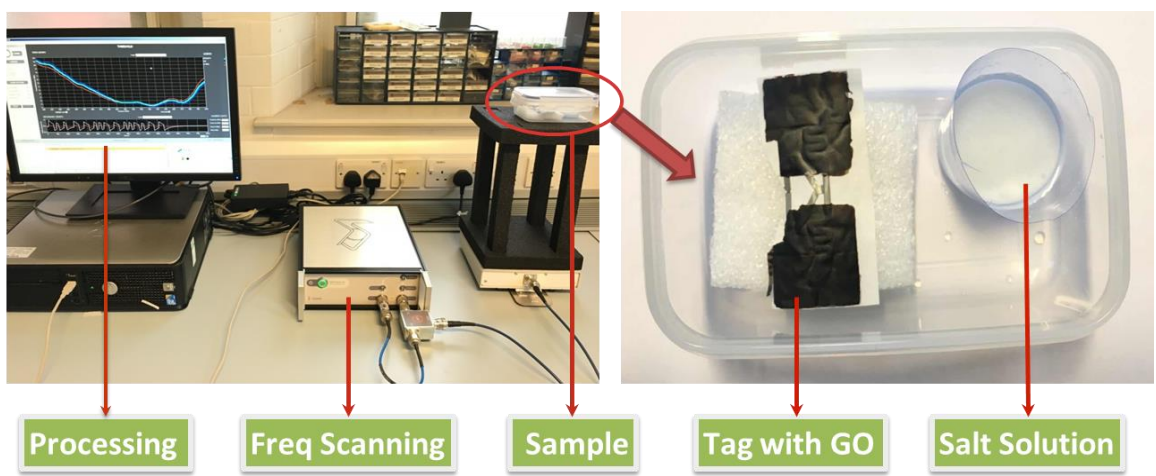

Fig. S4. Experimental setup for wireless RFID GO humidity sensing system.

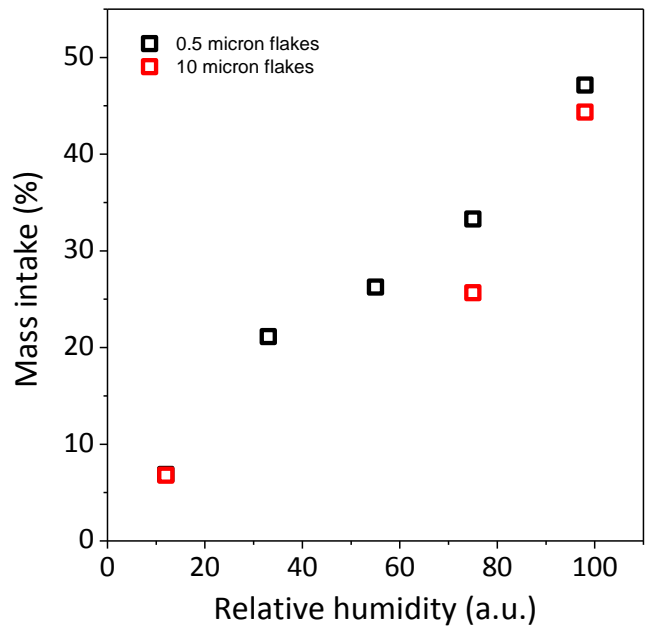

Fig. S5. Water uptake in GO samples. Relative increase in weight of GO prepared from different flake sizes exposed to different humidity.

Fig. S5 shows the measurement results of water uptake of GO with two different flake sizes (0.5 μm and 10 μm) by monitoring the weight change of GO exposed to different humidity conditions.

GO samples were completely dried inside a glove box before exposing to different humidity for 5 days. Mass uptake monotonically increases with increasing humidity and it varies from ~ 5% to 50%.

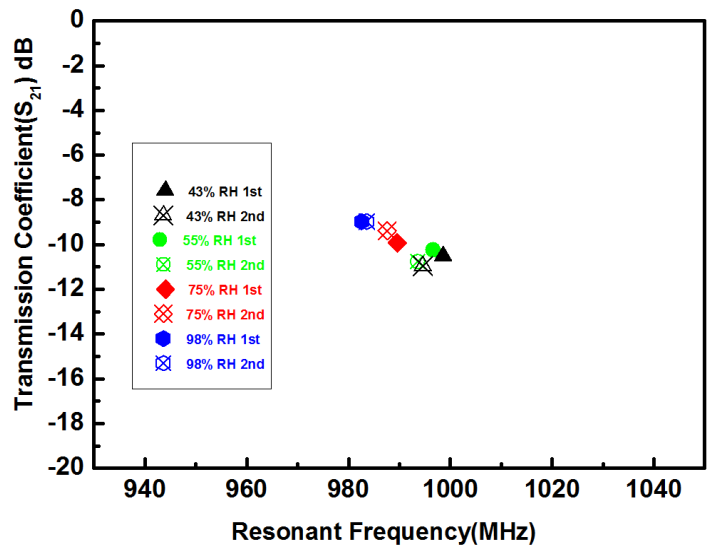

Fig. S6. Durability test of frequency response (x axis) and transmission response (y axis) of the GO coated resonator for various RH.

Fig. S6 shows the durability test of the resonator coated with GO for 43% RH to 98% RH. The first measurement was taken before drafting the manuscript in December 2015. The second measurement was made in January 2017. The time span is 13 months. All the measurement data collected were after 96 hours of humidity equilibration time. The x axis is the resonance frequency which indicates the frequency response to the GO coated resonator. The y axis is the transmission coefficient which indicates the propagation of the GO coated resonator. It can be seen that the measured results agree well with the previous data in both frequency response and the propagation level.

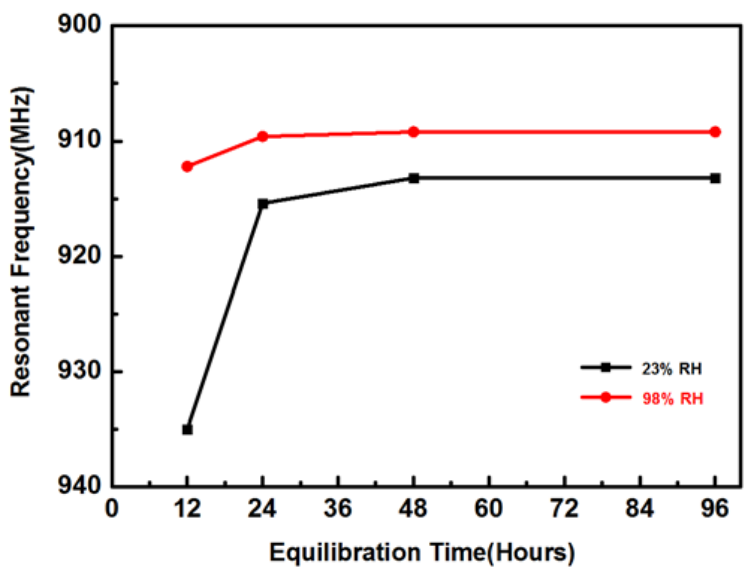

Fig. S7. Stability test of resonance frequency of GO coated sensor over equilibration time in room temperature (RH=23%) and with $K_2SO_4$, (RH=98%).

Fig. S7 shows the measured stability of the GO sensor. The resonance frequencies of GO coated sensor were measured with 23% RH and 98% RH, respectively. After 12 hours of equilibrium time, the resonance frequency still changes. The resonance frequency becomes stable and shifts no more after 48 hours.